\documentclass[12pt,preprint]{aastex}
\shorttitle{Blue Compact Dwarf Galaxy Mrk59}
\shortauthors{Thuan et al.}
\begin{document}

\title{{\sl FUSE} observations of the Blue Compact Dwarf Galaxy 
Mrk 59\footnote{Based on observations made with the NASA-CNES-CSA
{\sl Far Ultraviolet Spectroscopic Explorer}. {\sl FUSE} is operated
for NASA by the Johns Hopkins University under NASA contract NAS5-32985.}}
\author{Trinh X. Thuan}
\affil{Astronomy Department, University of Virginia,
    Charlottesville, VA 22903}
\email{txt@virginia.edu}

\author{Alain Lecavelier des Etangs}
\affil{Institut d'Astrophysique de Paris, CNRS, 98 bis bld Arago, 
F-75014 Paris, France}
\email{lecaveli@iap.fr}

\and

\author{Yuri I. Izotov}
\affil{Main Astronomical Observatory, National Academy of Sciences of Ukraine, 03680, Kyiv, Ukraine}
\email{izotov@mao.kiev.ua}

\begin{abstract}

New {\sl FUSE} far-UV spectroscopy of the nearby metal-deficient ($Z_\odot$/8) 
cometary Blue Compact Dwarf (BCD) galaxy Markarian (Mrk) 59 is discussed.
The data are used to investigate element abundances in its 
interstellar medium.
The H {\sc i} absorption 
lines are characterized by narrow cores which are interstellar 
in origin and by broad wings which are stellar in origin. The 
mean interstellar H {\sc i} column density is $\sim$ 
7 $\times$ 10$^{20}$ cm$^{-2}$ in Mrk 59. No H$_2$ lines are seen and 
$N$(H$_2$) is  $\la$ 10$^{15}$ cm$^{-2}$ at the 10 $\sigma$ level.
The lack of diffuse H$_2$ is due to 
the combined effect of a strong UV radiation field which destroys the 
 H$_2$ molecules and a low metallicity which leads to a scarcity of dust 
grains necessary for H$_2$ formation.
 P-Cygni profiles of the S {\sc vi} 933.4, 
944.5 \AA\ and O {\sc vi} 1031.9, 1037.6 \AA\ lines are seen, indicating 
the presence of very hot O stars and 
a stellar wind terminal velocity of $\sim$ 1000 km s$^{-1}$.
By fitting the line profiles with multiple components having each  
a velocity dispersion $b$ = 7 km s$^{-1}$ and spanning a radial velocity 
range of 100 km s$^{-1}$, some of which can be saturated, we derive 
heavy element abundances in the neutral gas. 
We find log $N$(O {\sc i})/$N$(H {\sc i})
 = --5.0$\pm$0.3 or [O {\sc i}/H~{\sc i}] = --1.5 for the neutral gas, 
about a factor of 10 below the oxygen abundance of the supergiant 
H {\sc ii} region, implying self-enrichment of the latter.

\end{abstract}



\section{Introduction}

The Blue Compact Dwarf (BCD) galaxy Markarian 59 (Mrk 59) $\equiv$ I Zw 49 
belongs to the 
class of cometary BCDs defined by Loose \& Thuan (1985) as 
characterized by an intense starburst at the end of an elongated 
low surface brightness (LSB) stellar 
body. In the case of Mrk 59, the elongated body is named NGC 4861. 
Arp (1966) describes NGC 4861 in his Atlas of Peculiar Galaxies as ``an 
object with irregular clumps, resolved into knots with a very bright knot 
(diameter = 1 kpc) at the southeastern end''.  
The knots are in fact a chain of 
H {\sc ii} regions, resulting probably from propagating star formation along 
the galaxy's elongated body and ending with the high-surface-brightness 
supergiant H {\sc ii} region at the southeastern end (Noeske et al. 2000). 
Dinerstein \& Shields (1986) first detected broad
Wolf-Rayet stellar features in Mrk 59 indicating the presence of late nitrogen
and early carbon Wolf-Rayet stars. Guseva, Izotov \& Thuan (2000) 
using the spectroscopic observations of Izotov, Thuan \& Lipovetsky (1997) 
found that several dozens of Wolf-Rayet stars are present. 
Noeske et al. (2000) used deep ground-based spectrophotometric observations  
of the supergiant H {\sc ii} region to 
derive an Oxygen abundance log O/H = --4.011 $\pm$ 0.003 
($Z_\odot$/8), typical of BCDs.
O abundances were also derived for two other emission knots along the elongated 
body and were found to be the same within the errors. The small scatter in metallicity 
along the major axis of Mrk 59 ($\sim$ 0.2 dex)
 suggests that the mixing of elements in the ionized gas 
has been efficient on a spatial 
scale of several kpc. 
Wilcots, Lehman \& Miller (1996) and Thuan, L\'evrier \& Hibbard (2001) 
have used the VLA to map NGC 4861 and 
Mrk 59 in the 21 cm line. With a beam size of 15\arcsec\ (780 pc), the latter 
authors found a very clumpy interstellar medium (ISM) 
 with H {\sc i} column densities ranging from a few 10$^{19}$ to a few 
10$^{21}$ cm$^{-2}$. 
Despite irregularities, the H {\sc i} velocity field of NGC 4861 resembles that 
of a rotating disk seen almost edge-on.
Spectral population synthesis in combination with color-magnitude diagrams 
and color profiles give a most probable age of $\sim$ 2 Gyr for the LSB 
elongated body, with an upper limit of $\sim$ 4 Gyr (Noeske et al. 2000), 
considerably smaller than the typical age (5 Gyr or greater) of the 
underlying stellar population in BCDs of other types, 
and making Mrk 59 a relatively young galaxy.

Because of these interesting properties and its brightness in the 
far-ultraviolet (Kinney et al. (1993) list an {\sl IUE} 
$\lambda$1432 -- 1532 \AA\ flux 
of $\sim$ 10$^{-13}$ erg s$^{-1}$ cm$^{-2}$ \AA$^{-1}$), Mrk 59 is 
a prime target in our {\sl Far Ultraviolet Spectroscopic Explorer (FUSE)}
 studies of BCDs. In particular, we 
wish to study the molecular hydrogen content and heavy element abundances in
its interstellar medium (ISM).  
Thuan et al. (2001) obtained 
a H {\sc i} heliocentric velocity of 847 km s$^{-1}$. 
Using the Virgocentric infall 
model of Schechter (1980) with parameters $\gamma$ = 2, $v$(Virgo) = 
976 km s$^{-1}$, $w$$_\odot$ = 220 km s$^{-1}$ and $D$(Virgo) = 15.9 Mpc, then 
the distance of Mrk 59 is 10.7 Mpc. At this distance, 1\arcsec\ corresponds to 
52 pc. Mrk 59 has $m_B$ = 12.64 (Noeske et al. 2000), so that $M_B$ = --17.51.

We describe the {\sl FUSE} observations in section 2. In section 3, 
we derive the Hydrogen 
column density by fitting the profiles of the H {\sc i} Lyman series 
lines. In section 4, we set upper limits on the amount of diffuse H$_2$. 
In section 5, we  
derive the interstellar ionic abundances by fitting the profiles 
of the metallic absorption lines with multiple components,
and compare the heavy element abundances in the neutral and ionized gas. 
In section 6, we discuss the O {\sc vi} and S {\sc vi} stellar lines 
with their P-Cygni profiles, and derive a terminal stellar wind velocity. 
We summarize our findings in section 7.         

\section{Observations}

Mrk 59 ($\alpha_{2000}$ = 12$^{\rm h}$59$^{\rm m}$00\fs3, $\delta_{2000}$ = 
34\arcdeg50\arcmin42\farcs8; $l$$^{\rm II}$ = 111\fdg54, 
$b$$^{\rm II}$ = 82\fdg12)
was observed during 7,865 s on 2000, January 11 with {\sl FUSE} 
(Moos et al. 2000). The LWRS large entrance 
aperture (30\arcsec $\times$ 30\arcsec) was used, so that all of Mrk 59 
which is $\sim$ 20\arcsec\ across (Noeske et al. 2000) is included within it.
Flux was recorded in both long wavelength (LiF, $\sim$ 1000 -- 1187 \AA) 
and both short wavelength (SiC, $\sim$ 900 -- 1100 \AA) channels. 
The pipeline version 1.5 has been used to process the data.
The spectral resolution is defined by both the instrument and the size of the 
galaxy. We find it to be about 10,000 with a signal-to-noise ratio (S/N)
 of $\sim$ 10 per resolution element. This a relatively high S/N for this 
type of object as observed by {\sl FUSE} (compare for example 
the spectrum of Mrk 59 in Figure 1 with that 
of the BCD I Zw 18 obtained by Vidal-Madjar et al. 2000).

The {\sl FUSE} 
spectrum shows two absorption line systems at two different radial 
velocities: one system of H$_2$ lines is at --250 km s$^{-1}$ and the other is 
at 600 km s$^{-1}$. 
Since the  H$_2$ lines arise in the 
Milky Way, there is a systematic wavelength shift of --250 km s$^{-1}$
over the whole spectrum. Taking into account that shift, the other 
absorption line system corresponds to a velocity of 850 km s$^{-1}$ which 
matches well the H {\sc i} heliocentric velocity of 847 km s$^{-1}$ of Mrk 59 
(Thuan et al. 2001). A wavelength blueshift of 100 km s$^{-1}$ has also been 
seen in the {\sl FUSE} spectrum of the BCD I Zw 18 by 
Vidal-Madjar et al. (2000).
This systematic wavelength shift is probably due to the 
preliminary wavelength calibration of {\sl FUSE} and a slight 
offset of the target from the center of the slit. 

\begin{figure}
\figurenum{1}
\epsscale{0.9}
\plotone{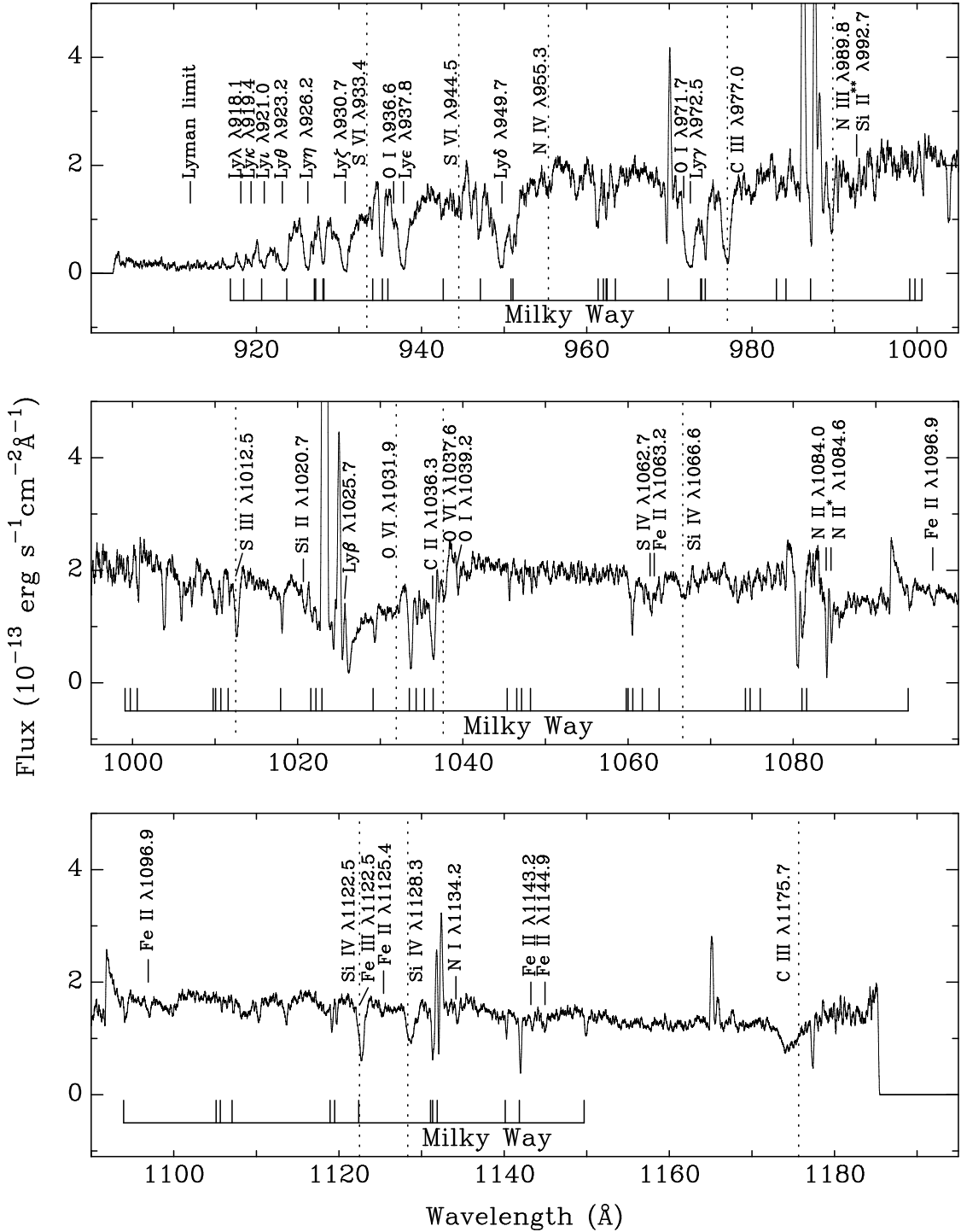}
\caption{\label{Fig1} Rest-frame {\sl FUSE} spectrum of Mrk 59. 
The data has been smoothed by a 11-point box-car. Prominent interstellar 
absorption lines are indicated. The lines arising in Mrk 59 are marked on top 
and those in the Milky Way at bottom. Particularly prominent in Mrk 59 are the 
H {\sc i} Lyman series lines. There are also strong interstellar absorption 
lines from the C {\sc ii}, C {\sc iii}, N {\sc i}, N {\sc ii}, O {\sc i}, 
S {\sc iii}, S {\sc iv}, Fe {\sc ii} and Fe {\sc iii} lines. The rest 
wavelenghts of the stellar lines in Mrk 59 are shown by dotted lines. 
The S {\sc vi}, O {\sc vi} and C {\sc iii} $\lambda$1175.7 lines show P-Cygni 
profiles.}
\end{figure}

Figure \ref{Fig1} shows the entire {\sl FUSE} spectrum 
shifted to the rest-frame of Mrk 59. We have marked 
the most prominent interstellar absorption 
lines: the H {\sc i} Lyman series and absorption 
lines from other atoms and ions such as C {\sc ii}, C {\sc iii}, N {\sc i}, 
N {\sc ii}, N {\sc iii}, O {\sc i}, Si {\sc ii}, S {\sc iii}, 
S {\sc iv}, Fe {\sc ii} and Fe {\sc iii}. 
Several stellar features such as the Si {\sc iv} and S {\sc vi} lines 
have been also detected and are marked by dotted lines.
There are no evident emission features except for the blue emission in the 
P-Cygni profiles of the stellar O {\sc vi} and S {\sc vi} lines. 
We have also marked the rest wavelengths of the 
interstellar absorption lines arising in the Galaxy 
at the bottom of each panel in Figure \ref{Fig1}.

\section{Narrow cores and broad wings of the H {\sc i} Lyman series   
line profiles: interstellar and  
stellar absorption}

The relatively high S/N of the {\sl FUSE} spectrum of Mrk 59 gives us 
an unique 
opportunity to study the absorption lines from the nearly entire Lyman 
series of Hydrogen from Ly $\beta$ to Ly $\lambda$ (H {\sc i} 11), 
except for Ly $\alpha$ which is out of the wavelength 
range covered by {\sl FUSE}. 
The foreground Galactic H {\sc i} column density 
in the direction of Mrk 59 is  2.8 $\times$ 10$^{20}$ cm$^{-2}$ (Heiles 1975).
Adopting a smooth continuum, we have used Voigt profiles
to fit the observed H {\sc i} Lyman series lines.
The fitting routine used 
was a modified version of the Owens profile fitting procedure
developed by Martin Lemoine and the {\sl FUSE} French team.
This program returns the most likely values of many free parameters like
the Doppler widths and column densities by a $\chi^2$ minimization of
the difference between the observed and computed profiles.
The latest version of this program is particularly suited to 
the characteristics of {\sl FUSE} spectra. For example,  it allows for a 
variation of the background level, a double gaussian point spread
function, and a shift of the wavelength scale. These are taken as free
parameters which are dependent on the wavelength region and determined 
by the $\chi^2$ minimization. A special version has been developed
to fit the absorption lines in the particular case of galaxies
where the observed spectrum is the sum of the fluxes
coming from thousands of stars through different absorbers with varying 
column densities. 

\begin{figure}
\figurenum{2}
\epsscale{1.0}
\plotone{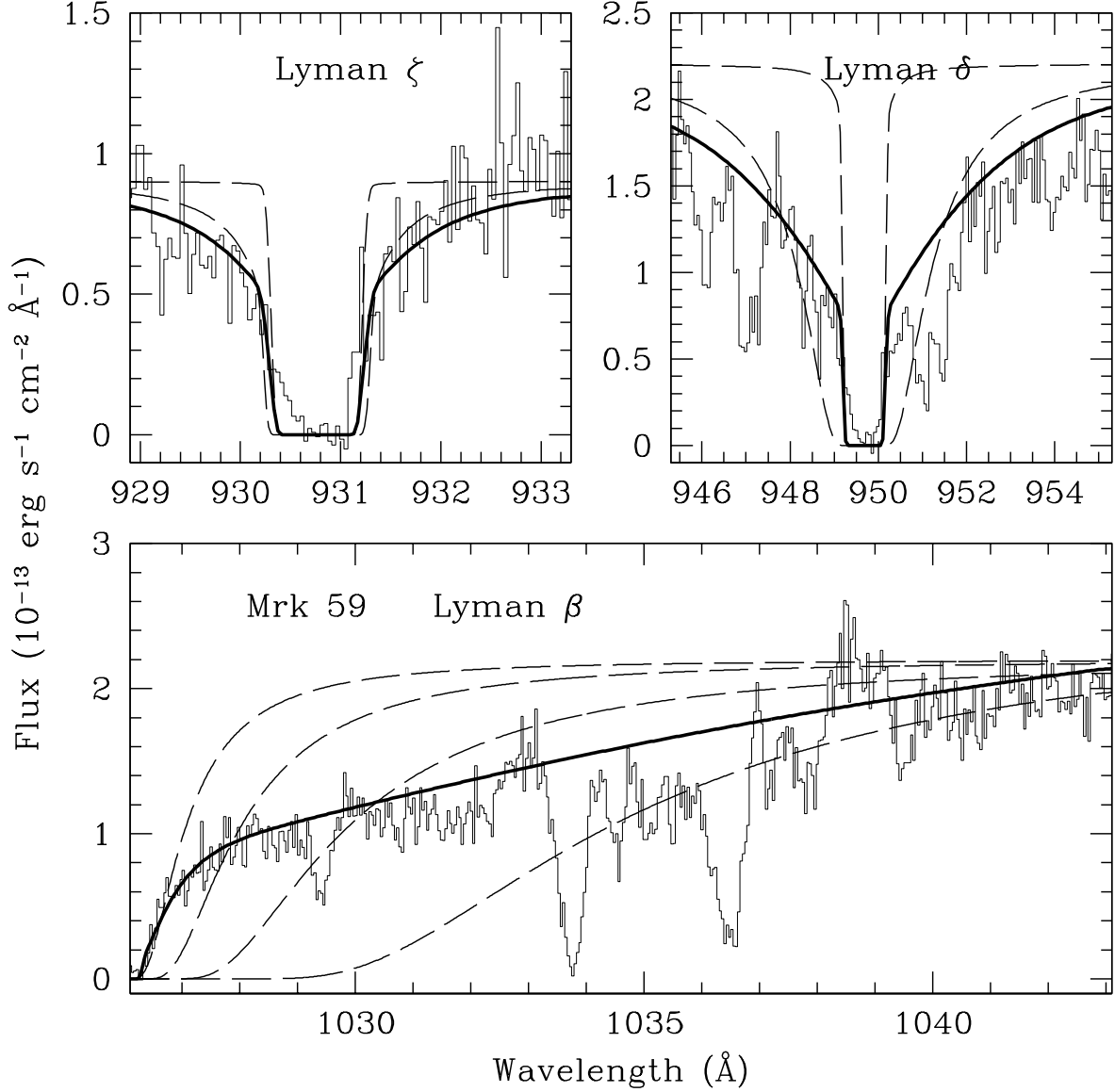}
\caption{\label{Fig2} Profile fitting of 3 H {\sc i} Lyman series 
lines. The solid lines show the best fit, with the distribution of H {\sc i} 
column densities shown in Fig. \ref{Fig3}. The dashed lines show fits with a 
single column density of respectively 10$^{21}$ and 5 $\times$ 10$^{22}$
cm$^{-2}$ in the case of the Lyman $\zeta$ and $\delta$ lines and
10$^{21}$, 3 $\times$ 10$^{21}$, 10$^{22}$ and 5 $\times$ 10$^{22}$
cm$^{-2}$ in the case of the red wing of the Lyman $\beta$ line. It is
evident that a single column density does not provide a good fit 
to the entire profile. We argue that the cores of the lines are 
interstellar in origin while the broad wings are stellar in origin.}
\end{figure}

We found, to our surprise,
that the shapes of the H {\sc i} Lyman series lines 
 cannot be reproduced with profiles 
characterized by a single H {\sc i} column density. 
Figure \ref{Fig2} illustrates the problem.
The upper-left panel shows the Lyman~$\zeta$ line. Damping wings are clearly 
visible, signaling a very large H {\sc i} column density $N$( H {\sc i}) $\sim$ 
$5\times 10^{22}$~cm$^{-2}$. The fit corresponding to that column density  is 
shown by the lower dashed line.  
However, this high column density is not consistent with the shape of the
other H {\sc i} lines. For example, the upper-right panel shows 
the H {\sc i} Lyman~$\delta$ line.
Although damping wings are also present, the core of the Lyman~$\delta$ line 
is thin and cannot be fitted with a model profile with  
the column density needed to fit the Lyman~$\zeta$ line (lower dashed line),
the predicted core being too wide.
The thin core implies a smaller column density, of the order of 
$10^{21}$~cm$^{-2}$ (upper dashed line), 
the expected value for a line of sight going through the ISM of the 
whole galaxy. 
Thus the problem is the following: assuming that the H {\sc i} Lyman series 
absorption lines are of interstellar origin, then at least two types 
of H {\sc i} column densities are needed, one type 
with $N$ (H {\sc i}) $\sim$ $10^{21}$~cm$^{-2}$ and the other with 
$N$ (H {\sc i}) $\sim$ 5 $\times$ $10^{22}$~cm$^{-2}$. A
bimodal distribution of  $N$ (H {\sc i}) is also needed in the fitting of   
the Lyman~$\beta$ line shown in  
the bottom panel of Figure \ref{Fig2}. We can only fit 
its red wing as 
the blue wing is contaminated by the Earth's airglow. 
The dashed lines show respectively, from bottom to top, 
model profiles for H {\sc i} column 
densities of $5\times 10^{22}$, $10^{22}$, $3\times 10^{21}$, and 
$10^{21}$~cm$^{-2}$.
None of these profiles is able to fit the observed shape of the line.
Again, a very large H {\sc i} column density
 of the order of $10^{23}$ cm$^{-2}$ 
is needed to explain the broad 
damping wings, whereas its core is very narrow
requiring a column density of the order of $10^{21}$~cm$^{-2}$.

A first interpretation of the bimodality of $N$(H {\sc i}) 
is that the absorbing interstellar 
medium in front of the far-UV-bright stars is very inhomogeneous.
Since the observed spectrum is the  
sum of spectra of individual stars located behind H {\sc i} clouds with 
different column densities, and there are more than 10$^3$ lines of sight to 
stars emitting in the far-UV over the $\sim$ 20\arcsec\ wide central region 
of Mrk 59, these may go through H {\sc i} clouds with widely 
different column densities.
To test that hypothesis in a more rigorous manner, 
we have performed a fit of the H {\sc i} Lyman series lines by 
considering various mixtures of 23 different H {\sc i} column
densities which span nearly 3 orders of magnitude, 
from a few $10^{20}$ to $10^{23}$~cm$^{-2}$. 
The resulting fit is found by $\chi^2$ 
minimization of the difference between the computed and observed profiles 
of the H {\sc i} lines, from
Lyman $\beta$ to Lyman $\eta$, when varying the fractions of 
the contributing column densities. That fit is shown by  
solid lines for the Ly $\beta$, Ly $\delta$ and Ly $\zeta$ lines in 
Figure 2.  
The corresponding H {\sc i} column density distribution is shown in 
Figure 3. As expected from the consideration 
of individual lines, the $N$ (H {\sc i})
distribution is bimodal, 
with one peak near 5 $\times$ $10^{20}$~cm$^{-2}$ and the other 
near $10^{23}$~cm$^{-2}$.
As discussed before, the high H {\sc i} column densities are
required to fit the damping wings of the H {\sc i} lines 
further up in the Lyman 
series, particularly from the Ly $\delta$ to the Ly $\lambda$ lines. 
On the other hand,
the core of the Lyman lines further down the series are considerably 
narrower. For example, the Lyman~$\beta$ line
has a core width of less than 1 \AA, corresponding to a column density 
below $10^{21}$~cm$^{-2}$.  About 1/3 of the Mrk~59 far-UV 
flux is seen through H {\sc i} clouds with a ``normal'' column density 
while about 2/3 of the flux is absorbed
by H {\sc i} with very high column densities.

There are several problems with 
attributing the very high $N$(H {\sc i}) to interstellar clouds. First, 
H {\sc i} interstellar column densities 
as high as $10^{23}$~cm$^{-2}$ are not seen. The highest known 
H {\sc i} column density known thus far in a BCD is $7\times 10^{21}$ cm$^{-2}$
in SBS 0335--052 (Thuan \& Izotov 1997). 
With a cloud of column density $\sim$ $10^{23}$~cm$^{-2}$ 
covering the whole galaxy, the core of the Lyman~$\beta$ line would have been 
several \AA\ wide, which is not observed.
Second, if the high column density is due to an interstellar cloud 
covering about $\sim$ 2/3 of the area of 
Mrk~59, the associated dust would strongly absorb the far-UV 
photons, which is not the case. 

We thus need to find other explanations for the broad damping wings 
responsible for the high derived column densities. One source of line 
broadening is wavelength smearing effects due to the extended nature of the 
source. However, we do not believe those to be the main cause of the broad 
H {\sc i} wings because of several reasons. First,
they fall far short from accounting for the very 
large widths of the damping wings.  Indeed the size of the {\sl FUSE} aperture 
(30 arcsec) and the size of the observed galaxy (less than 20 arcsec) give 
a wavelength smearing of $\pm$30 km s$^{-1}$. This is significanty below the
half width at half-maximum of  
the Fe {\sc iii} line which is larger than 100 km s$^{-1}$, and more 
than one order of magnitude lower than the half-widths of the wings 
of the H {\sc i} lines, 
observed to be $\sim$ 1000 km s$^{-1}$ for the Ly$\delta$
and Ly$\gamma$ lines, and above 4000 km s$^{-1}$ for the Ly$\beta$ line. 
Second, many other lines are very narrow. For instance,
the Galactic H$_2$ absorption lines have full widths at half-maximum (FWHM) 
less than 30 km s$^{-1}$. If the broad wings are due solely to 
wavelength smearing, then all lines should show them, not just hydrogen lines.
Third, the hydrogen lines in the {\sl FUSE} spectrum of the  
BCD I Zw 18 which is comparable in angular size ($\sim$ 20$\arcsec$) and 
was obtained through the same aperture 
(Vidal-Madjar et al. 2000) do not show the broad wings seen in Mrk 59. Hence, 
the latter cannot be attributed to instrumental effects. While wavelength 
smearing effects due to the extended nature of the source may indeed give 
a measured velocity dispersion $b$  
larger than the true value,
they cannot be mainly responsible for the broad wings of the H {\sc i} lines.

Given that the broad wings of the higher order Lyman lines are not caused by  
instrumental effects, what physical effect may be responsible for
them? The possibility that they are due to 
H {\sc i} cloud motions is unlikely because the lines have  
symmetric profiles and their large widths 
correspond to velocities of several hundred to more than 
a thousand km s$^{-1}$. Such high velocities are not evident either in the 
single-dish H {\sc i} profile of Mrk 59 (Thuan \& Martin 1981) or in its 
H {\sc i} VLA map (Thuan et al. 2001). 
A more plausible explanation is that the wide damping wings
are not interstellar in origin, but stellar. 
Many of the broad absorption features in the {\sl FUSE} spectrum
of Mrk 59 (Fig. 1) such as the S {\sc vi} 933.4, 944.5 \AA, 
O {\sc vi} 1031.9, 1037.6 \AA\ lines
with P Cygni profiles, the broad Si {\sc iv} 1128.3 \AA, 
N {\sc iv} 955.3 \AA\ and blue-shifted 
C {\sc iii} 1175.7 \AA\ lines are undoubtedly formed in atmospheres of
massive stars with spectral types ranging 
from the earliest O to early B types. Some
other broad lines such as the H {\sc i} Lyman series lines discussed before, 
the C {\sc iii} 977.0 \AA, N {\sc iii} 989.8 \AA, S {\sc iii} 1012.4 \AA\
and Fe {\sc iii} 1122.5 \AA\ 
lines can be formed
both in the interstellar medium and in the atmospheres of stars.
Examination of far-UV spectra of early
B stars in the effective temperature range 20000 -- 25000 K 
based on Kurucz (1991)'s models does indeed 
reveal hydrogen and ion lines with considerable wings.
H {\sc i} Lyman lines with broad wings are also seen in {\sl FUSE}
spectra of B stars (Fremat et al. 2001). Thus we conclude that 
the broad wings of the above lines in Mrk 59  
arise in the photospheres of the numerous B stars in the BCD,
while their narrower cores are caused by interstellar absorption.
 
We point out however that, besides the B stars, 
there is also a population of more massive early O 
stars in Mrk 59. This is evidenced by the detection of the 
S {\sc vi} 1031.9, 1037.6 \AA\  and O {\sc vi} 1031.9, 1037.6 \AA\ lines 
in the {\sl FUSE} spectrum (section 6). 
These lines are only present in the most massive stars. 
Additional support for the presence of a population of early
O stars comes from the optical spectrum of Mrk 59 (e.g., Izotov et al. 1997; 
Guseva et al. 2000) where
strong emission lines are present. The equivalent
width ($EW$) of the H$\beta$ emission line of $\sim$ 150 \AA\ puts 
strong constraints on the age of the starburst. 
In the case of an instantaneous burst model, 
the observed $EW$(H$\beta$) gives a burst age of only $\sim$ 4 Myr, 
approximately the lifetime of a 60 $M_\odot$ star (Schaerer \& Vacca 1998). 
In the case of continuous star
formation, the observed $EW$(H$\beta$) can be reproduced by 
a stellar population constantly forming 
between 0.1 and 10 Myr. In this case, massive stars should also 
be present. Finally, evidence for the presence of very massive stars in Mrk 59 
comes from the detection of broad Wolf-Rayet features at 4650 \AA\
and 5808 \AA. According to Schaerer \& Vacca (1998), at the metallicity
of Mrk 59, Wolf-Rayet stars are descendants of main-sequence stars
with masses greater than 40 -- 50 $M_\odot$.

Thus, while stars from early O to early B types are responsible for  
the continuum distribution and 
its slow intensity decrease for wavelengths shorter than 950 \AA,
 the broad wings of the H {\sc i} Lyman series lines and some heavy element
lines originate in the photospheres of early B stars.
We note that the wings of the  H {\sc i} Ly$\beta$ line are not as 
broad in the BCD I Zw 18 so that both core and wings 
can be fitted with a 
single interstellar H {\sc i} column density (Vidal-Madjar et al. 2000).
 This can be  
due to I Zw 18 having a younger stellar population with predominantly hot 
O stars in its star-forming region. However, only the Ly$\beta$ line
has been fitted in I Zw 18, half the {\sl FUSE} wavelength range being 
not available in the Vidal-Madjar et al.' spectrum. This may not be 
enough to reveal the problem. 

By fitting the cores of the Lyman series lines, we obtain an  
interstellar H~{\sc i} column density in Mrk 59 of $N$(H {\sc i}) $\sim$ 
7 $\times$ 10$^{20}$~cm$^{-2}$. $N$(H {\sc i}) is estimated as the 
weighted mean of all column densities contributing to 
the first peak in the bimodal distribution of Figure 3, i.e 
all those with a value less than 10$^{22}$~cm$^{-2}$.

\section{Upper limits on the diffuse H$_2$ content of Mrk 59}

H$_2$ lines from the Milky Way are 
detected up to levels of at least J = 3 in the spectrum of Mrk 59, 
with column densities of a few $10^{20}$~cm$^{-2}$ (Fig. 1). 
However, no line of H$_2$ is seen at the radial velocity of the BCD
despite the fact that Mrk 59 has H {\sc i} column densities very similar to 
those in the Milky Way.

From a VLA H {\sc i} map of Mrk 59 (Thuan et al. 2001), we derive 
a velocity dispersion $b$ = 20 km s$^{-1}$ which we use 
to calculate upper limits to the H$_2$ column densities. 
Using Voigt profiles for the H$_2$ lines, 
we construct a series of synthesized spectra with varying H$_2$ 
column densities, compute the 
difference between each synthesized spectrum and the observed spectrum in 
nine Lyman bands (0--0 to 8--0), and calculate the increase in the 
$\chi$$^2$ of the fit to the observed spectrum. The upper limits given 
in Table \ref{Tab1} for the $N$(H$_2$) column densities in Mrk 59 
correspond to an increase of 
$\chi$$^2$ by a factor of 100, meaning that they are 10 $\sigma$ limits.
Because most of the H$_2$ molecules are in the J = 0 or J = 1 levels at 
temperatures typical of the interstellar medium, 
we conclude that the total column density of H$_2$ is $\la$
$10^{15}$~cm$^{-2}$.
Fig. \ref{Fig4} shows the expected H$_2$ lines if the column density of 
H$_2$ had been 10$^{15}$ cm$^{-2}$ for the J = 0, 1 and 2 levels.
It is clear that H$_2$ with such a column density in Mrk 59 would have 
been easily detected by {\sl FUSE}.
With a H~{\sc i} column density of $\sim 7\times 10^{20}$~cm$^{-2}$,
this implies that the ratio of 
H$_2$ to H {\sc i} is $\la 10^{-6}$  
in the absorbing clouds in front of Mrk 59. The corresponding 
average molecular fraction
$f$ = 2$N$(H$_2$)/($N$(H {\sc i})+2$N$(H$_2$)) is $\la 3 \times 10^{-6}$.
Adopting a diameter of 1 kpc for Mrk 59 gives 
an upper limit of $\sim$ 12 $M_\odot$ for the total mass of diffuse H$_2$.

This is a very stringent 
upper limit. Only one BCD, I Zw 18, has been searched previously by {\sl FUSE} 
for diffuse H$_2$ (Vidal-Madjar et al. 2000). These authors also established 
an upper limit for the diffuse H$_2$ column density of $\sim 
10^{15}$~cm$^{-2}$,
corresponding to $M$(H$_2$) $\la$ 30 $M_\odot$. Thus our H$_2$ upper limits 
for Mrk 59 confirm and strengthen the results of Vidal-Madjar et al.
 Of course, these limits concern only 
diffuse H$_2$ along the line of sights to 
the several thousands of far-UV bright massive stars in Mrk 59. Our {\sl FUSE}
observations are not 
sensitive to clumpy H$_2$ in very compact 
star-forming regions. High dust extinction in these 
regions would prevent the far-UV photons of the massive young stars from 
escaping. The dust 
distribution being clumpy, the far-UV photons go out preferentially
along lines of sight that are devoid of dust. This is 
deduced from the observation that the extinction derived 
from UV spectra is significantly smaller than the value derived from 
the Balmer decrement as measured from the optical spectra of the same object. 
The UV light preferentially emerges through windows of low optical depth 
(Fanelli, O'Connell \& Thuan 1988). This implies that, in some regions,
 the dust is 
efficiently destroyed or removed from the vicinity of young massive 
O stars by the combined action of ionizing radiation, stellar winds, 
and supernovae. Since dust is needed for efficient formation of H$_2$ 
molecules,
its absence along particular lines of sight implies that these 
molecules will also be absent along those directions. 

Two possible formation mechanisms 
of H$_2$ without dust grains have been proposed. The first mechanism is 
via the formation 
of the negative ion H$^-$ (H + e $\rightarrow$ H$^-$ + e), followed by the 
associative reaction  H$^-$ + H $\rightarrow$ H$_2$ + e, where the electrons 
come from the photoionization of carbon (Jenkins \& Peimbert 1997). 
The second mechanism consists of the 
radiative association of H and H$^+$ to form H$_2^+$, which then results in 
H$_2$ by the capture of an electron. However these two mechanisms are 
extremely inefficient under the temperatures (10$^2$ -- 10$^4$ K) and 
densities 
(10$^2$ -- 10$^4$ cm$^{-3}$) prevailing in the ISM of Mrk 59 and will not be 
discussed further. We shall thus only consider the case where H$_2$ is formed 
on the surface of grains. 
   
In that case, to detect H$_2$ molecules, 
we need to probe dusty regions. Near-infrared (NIR) and mid-infrared 
(MIR) observations serve well that purpose  
since infrared photons suffer less extinction from dust and 
can escape more easily than UV photons.   
Indeed, H$_2$ emission lines are 
seen in the NIR emission-line 
spectra of some of these objects (Vanzi \& Rieke 1997 
and Vanzi et al. 2000). Most spectacular is the case of the BCD SBS 0335--052
($Z_\odot$/41), the second most metal-deficient BCD known after I Zw 18 
($Z_\odot$/50). 
Thuan, Sauvage \& Madden (1999) derived from {\sl ISO} 
$\lambda$5--17 $\micron$ 
observations an extinction $A_V$ $\sim$ 20 mag for SBS 0335--052. Such a 
large extinction implies that a significant fraction (as much as $\sim$ 75\%)
of the current star formation activity in SBS 0335--052 is hidden by dust 
whose total mass is $\sim$ 3000 $M_\odot$. Optical photons cannot 
get out of these very compact star-forming regions, much less UV photons.
 There is some evidence that the strengths of the 
 H$_2$ emission lines are correlated with the amount of dust present in these 
objects. Thus the line intensity ratio of the (1,0)S(1) H$_2$ line to 
the Br$\gamma$ H {\sc i} line in SBS 0335--052 is among the highest observed 
for BCDs. 

To summarize, H$_2$ molecules are probably present in Mrk 59, but their 
distribution is clumpy and most likely follows that of the dust. 
{\sl FUSE} does not see 
this H$_2$ because it selects out line of sights going through regions 
devoid of dust. 

We now check whether the absence of diffuse H$_2$ is reasonable, given the 
physical conditions in the H {\sc i} cloud surrounding Mrk 59. 
These conditions must be quite different from those in the Milky Way
since Mrk 59 has H {\sc i} column densities similar to 
those seen in the Milky Way for which 
diffuse H$_2$ is strongly detected while it is 
conspicuously absent in the BCD (Fig. 4).
We first explore the situation at the edge of the cloud. From the VLA map 
of Thuan et al. (2001), the angular radius of the H {\sc i} cloud is 
180$\arcsec$ or $R_0$ = 9.4 kpc. From the {\sl FUSE} spectra, 
the radiation flux 
at 1000 \AA\ at Earth is $\sim$ 2.3 $\times$ 10$^{-13}$ erg s$^{-1}$ cm$^{-2}$ 
\AA$^{-1}$. Correcting for the 
Galactic extinction $E(B-V)$ = 0.011 mag (Schlegel, Finkbeiner \& Davis 1998) 
with $A$(1000 \AA) = 5.742 $A_V$ = 0.2 mag (Cardelli, Clayton \& Mathis 1989) 
gives a flux at $R_0$ from the ionizing stars $F_{R_0}$ = 
3.6 $\times$ 10$^{-7}$ erg s$^{-1}$ cm$^{-2}$ \AA$^{-1}$.
 We have assumed the internal extinction in 
the far-UV to be negligible, as argued before.  
By solving the grain temperature equilibrium equation 
obtained by equating the radiation absorbed by grains to that emitted,
and adopting the approximations by Vidal-Madjar et al. (2000),
 we obtain a grain temperature at the edge of the H {\sc i} cloud  
of 11 ($a$/0.1 \micron)$^{-1/6}$ K, where 
$a$ is the radius of the grain assumed to be spherical. The grain temperature 
increases as $R^{-1/3}$ closer to Mrk 59, and at the edge of the star-forming 
region at a radius of 10$\arcsec$ or $R_0$ = 0.52 kpc, it is 29 
($a$/0.1 \micron)$^{-1/6}$ K. 
We conclude that the range of grain temperatures in the H {\sc i} cloud around Mrk 59 
is very similar to that in the diffuse interstellar medium of the Milky Way,
and that H$_2$ formation can indeed occur.

To calculate the amount of H$_2$ molecules formed on  
the surface of dust grains, we assume that the H$_2$ formation rate scales 
with metallicity, i.e. that its value in Mrk 59 is 1/8 the Galactic value 
since the metallicity of Mrk 59 is 1/8 that of the Sun.
Following the assumptions described by Vidal-Madjar et al. (2000), 
at equilibrium, when the
formation of H$_2$ on dust equals its destruction rate by UV photons, 
the molecular fraction $f$(H$_2$) =  2$n$(H$_2$) / [2$n$(H$_2$) + $n$(H)]
(where $n$(H$_2$) and $n$(H) denote the number densities of H$_2$ and H) 
in Mrk 59 is 
proportional to $Z_{\rm Mrk\,59}n{\rm (H)}/ I$, 
where $I$ is the flux of H$_2$  
dissociating photons. We can write $f$(H$_2$) as:
$$
f({\rm H}_2) = 1.6\times 10^{-34} 
\left(\frac{Z_{\rm Mrk\,59}}{Z_{\odot}}\right) 
 \left(\frac{F_{R_0}}{{\rm erg\,s}^{-1}{\rm cm}^{-1}{\rm \AA}^{-1}}\right)^{-1}
 \left(\frac{R_0}{{\rm kpc}}\right)^{-1}
\left(\frac{N({\rm HI})}{{\rm cm}^{-2}}\right).
$$
With the approprate numerical values for Mrk~59, we find
$f(\rm{H}_2)$ $\sim$ 6 $\times$ 10$^{-30}$ $N$(H {\sc i}). 
Thus the H$_2$ column density $N$(H$_2$) is 
$\sim$ 3 $\times$ 10$^{-30}$ $N$(H {\sc i})$^2$.
For $N$(H {\sc i}) equal to 7 $\times$ 10$^{20}$ cm$^{-2}$, 
the expected $N$(H$_2$) is $\sim$ 2 $\times$ 10$^{12}$~cm$^{-2}$, nearly 
three orders of magnitude below and consistent with our 
established upper limit. As compared to the Milky Way, the low column density 
of diffuse H$_2$ in Mrk 59 
is due to the combined effects of a large UV flux which destroys the 
H$_2$ molecules and of a low metallicity resulting in 
a scarcity of dust grains on which to form them.

\section{Heavy element abundances}    
\label{Heavy element abundances}

We next derive heavy element abundances by fitting the profiles of 
the metal absorption lines. The line profile fitting method is superior to 
a simple curve-of-growth analysis because it allows to simultaneously 
fit the lines from different heavy elements in Mrk 59 and also the H$_2$ lines 
in the Milky Way. This permits in turn to constrain in a self-consistent manner
 several parameters such as 
the $b$ parameter, the instrumental profile and the background 
level. A curve-of-growth method does not allow easily for such an analysis.

In performing the line profile fitting, we have considered two cases. 
The first case is probably less realistic, but gives lower limits to 
the abundances in the interstellar medium of Mrk 59. 
It assumes that there is a single 
velocity component along the line of sight to the BCD. Consequently, lines 
that are broader than the point spread function and do not go down to 
zero intensity level are supposed to be not saturated.
The second case considers multiple velocity 
components along the line of sight, some of which may have saturated profiles. 
It is probably closer to the true situation.  

\subsection{Profile fitting with a single velocity component}      

\begin{figure}
\figurenum{3}
\epsscale{1.0}
\plotone{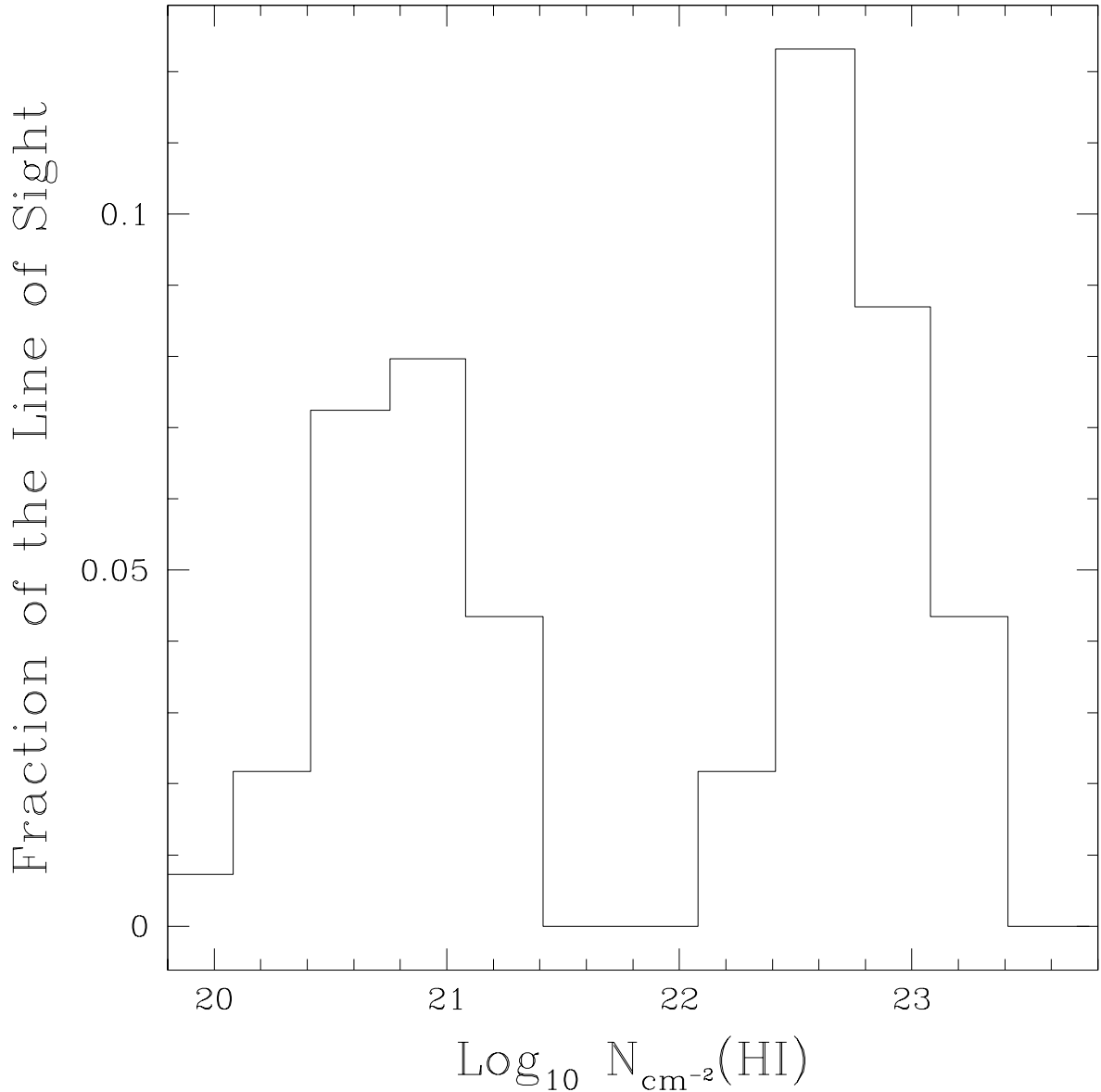}
\caption{\label{Fig3} Distribution of H {\sc i} column densities 
needed to fit the absorption profiles of some H {\sc i} Lyman series lines (see 
Fig. \ref{Fig2}). The distribution is bimodal: $\sim$ \onethird\ of the lines of sight to the far-UV bright sources have a H {\sc i} column density of a few
10$^{20}$ cm$^{-2}$, while $\sim$ \twothirds\ have a H {\sc i} column density of $\sim$ 10$^{23}$ cm$^{-2}$. We argue that the first peak is due to interstellar 
clouds while the second peak is due to stellar photospheres.}
\end{figure}

\begin{figure}
\figurenum{4}
\epsscale{0.8}
\plotone{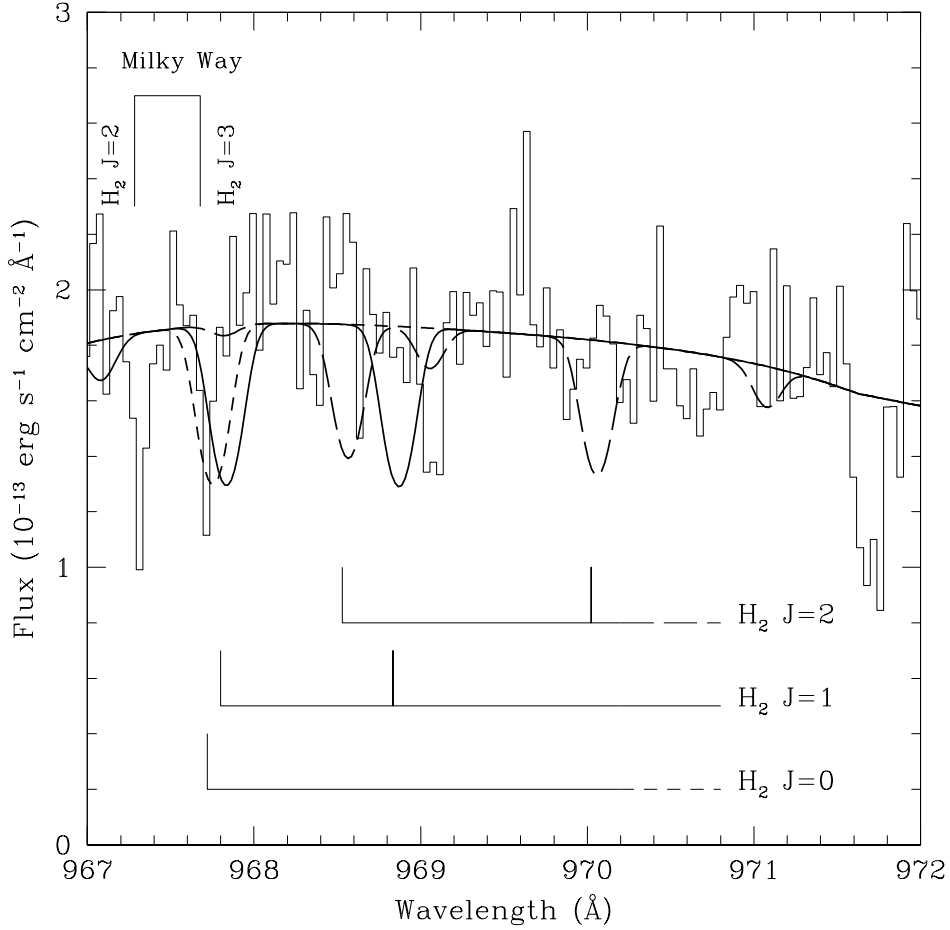}
\caption{\label{Fig4} Plot of the expected H$_2$ lines from the 
Werner series (0-2) if H$_2$ were present in Mrk~59 at the 10 $\sigma$ level.
 Molecular hydrogen lines from the Milky Way which has a H {\sc i} column 
density comparable to Mrk 59 are easily 
detected and marked here for the J = 2 and J = 3 levels. 
The spectral region shown is only one out of a total of  
18 regions in the spectrum used to constrain the upper limit of H$_2$. 
The H$_2$ line profiles corresponding to a H$_2$ column density of 10$^{15}$ 
cm$^{-2}$ are shown 
for the J = 0 (long-dashed line), J = 1 (solid line) and J = 2 (short-dashed line) levels.}
\end{figure}

We consider first the single interstellar velocity component case.
Several absorption lines of neutral and singly, doubly 
and triply ionized ions of heavy elements are detected in the {\sl FUSE} 
spectrum of Mrk 59. The most prominent ones are the 
C {\sc ii} 1036.3 \AA, C {\sc iii} 977 \AA,
N {\sc i} 1135.0 \AA,
N {\sc ii} 1084.0 \AA, O {\sc i} 1039.2 \AA,  S {\sc iii} 1012.5 \AA,   
S {\sc iv} 1062.7 \AA, Fe {\sc ii} 1144.9 \AA\ 
and Fe {\sc iii} 1122.5 \AA\ lines. They are are marked in Figure \ref{Fig1}
by solid lines.  
Some of the lines, for example the 
Fe {\sc iii} 1122.5 \AA\ and S {\sc iii} 1012.5 \AA\ lines, are very broad.
A single Voigt interstellar profile fit to these lines would give 
unreasonably large 
Doppler widths, $b$ $\ga$ 100~km~s$^{-1}$ (Table 2). The large widths 
suggest, as in the case of the H {\sc i} Lyman series lines, that 
these metal lines are not only interstellar in origin
 but also have some contamination by stellar absorption.
Indeed, we found that the profiles of these broad lines can be well fitted by 
using the bimodal distribution of column densities found for  
neutral hydrogen $N$(H {\sc i}) and shown in Figure \ref{Fig3},  
and by varying only a single parameter,
 the ratio of the species's column density to that of 
H {\sc i}, $N$(element)/$N$(H {\sc i}). 

For example, the fit to the S {\sc iii} 1012.5 \AA\ line profile 
based on the $N$(H {\sc i}) distribution in Fig. 3 is shown by 
a thick line in Fig. 5. No good 
fit can be obtained with a single ion column density and 
a reasonably low $b$ parameter, for example 42 km s$^{-1}$, 
as shown by the dashed 
lines in Fig. 5 for $N$(S {\sc iii}) equal 
to 8 $\times$ 10$^{14}$ and 
4 $\times$ 10$^{15}$ cm$^{-2}$. As in the case for the 
Lyman series H {\sc i} lines, none of these dashed profiles can account 
simultaneously for both the width and the depth of the observed line, the 
width of the line requiring high column densities, while its shallowness
requiring low column densities. The ratio which gives the 
best fit to the S {\sc iii} line profile is 
$N$(S {\sc iii})/$N$(H {\sc i}) = 8 $\times$ 10$^{-8}$. 
For the lines of other species like O {\sc i} 1039.2 \AA\ and N {\sc i}
1135.0 \AA, good fits can be obtained
with Voigt profiles with $b$ values between 20 and 40~km~s$^{-1}$.
The resulting column densities and Doppler widths $b$ 
along with their error bars  
are given in Table~\ref{Tab2} under the heading 
log $N_1$(species) for the most prominent 
interstellar absorption lines. The error bars of the column densities 
include the uncertainty in $b$. Only a lower limit to the column density 
is given for the C {\sc ii} 1036.3 \AA\ line. It is derived not from 
profile fitting, but from the equivalent width of the line as only 
its blue wing can be observed, its red wing being contaminated by 
Earth's airglow. 
A fit to the blue wing alone would give a considerably larger
value: $N$(C {\sc ii})$\gg 10^{17}$~cm$^{-2}$.

Taken at face value, the abundances of the metals relative to 
$N$(H {\sc i})  
are extremely low as compared to the solar values.
In particular, with log $N$(O {\sc i})/$N$(H {\sc i}) = --5.7 and
using the Meyer et al. (1998)' value for the Galactic ISM, we obtain
 [O {\sc i}/H {\sc i}] = --2.2. 
Thus, if the O {\sc i} line has its origin in the neutral gas,
the oxygen abundance in the H {\sc i} gas is $\sim$ 50 times lower  
than the oxygen abundance in the ionized gas, as determined from the optical 
emission line spectrum of the supergiant H {\sc ii} region in Mrk 59. It 
is $\sim$ 7 times lower than the oxygen abundance
in the star-forming region of I Zw 18, the most
metal-deficient BCD known. 
However, these very low abundances  
may be caused by the saturation of some lines, which we discuss next.

\subsection{Profile fitting with multiple velocity components}

If there is a single velocity component along the line of sight, then 
the fact that the lines, for instance the O {\sc i} and 
Fe {\sc ii} lines, do not go to zero intensity although they are
broader than the instrumental point spread function determined with the 
H$_2$ lines of the Milky Way, would argue for them not to be saturated.
However, because Mrk 59 is extended, we are observing the BCD 
along thousands of lines of sight through the {\sl FUSE} aperture, so  
the observed spectrum is the sum of many narrower spectra with different 
velocities. 
It can happen that some lines of sight have saturated absorption lines
with a small $b$ parameter,
 but because they have different radial velocities spread 
over several tens of km s$^{-1}$, 
the broad absorption line resulting from the sum of many narrow absorption 
lines does not go to zero intensity, and its  
width is larger than the point spread function. In that case, a 
single velocity component fit to the line profile  
would result in an overestimate of 
the $b$ parameter and in an underestimate of the column density.

To investigate this issue, we have calculated profiles resulting from the
addition of multiple lines of sight for all heavy element lines with 
good enough S/N. We adopt the simple model where  
the different lines of sight have radial velocities distributed 
uniformly between $v_{\rm Mrk 59} - {\Delta}v/2$ and $v_{\rm Mrk 59} + 
{\Delta}v/2$, where $v_{\rm Mrk 59}$ is the radial velocity of Mrk 59 
and ${\Delta}v$ is the spread in velocity due mostly 
to the velocity dispersion of the system of 
absorbing clouds along the multiple lines of sight, and also partly to the
wavelength smearing caused by the extension of Mrk 59 within the aperture.
We determine the best fit to several lines of 
O {\sc i}, N {\sc i} and Fe {\sc ii} simultaneously by  
varying $b$ (assumed to be the same for
all lines of sight) and ${\Delta}v$, so as to minimize the $\chi^2$ of the   
difference between the observed and computed profiles.
Because the spectrum of Mrk 59 shows  
several lines of the same element with different oscillator strengths,
the $b$ and ${\Delta}v$ parameters, which are independent of the chemical 
element considered, can be constrained reasonably well. 

The best solution was obtained for $b$ = 7$^{+13}_{-3}$ km s$^{-1}$ and 
${\Delta}v$ = 100$\pm$20 km s$^{-1}$, where the error bars are 2$\sigma$ 
limits. These values are physically quite reasonable. The $b$ value is 
similar to those found for individual absorbing clouds in the 
ISM of the Milky Way. As for the value of ${\Delta}v$, it is 
very close to the FWHM of 92 km s$^{-1}$ of 
the H {\sc i} profile (Thuan et al. 2001). 
However the 30$\arcsec$ {\sl FUSE} aperture does not sample all  
the neutral gas which extends over some 6$\arcmin$ in diameter. 
Thus the observed velocity spread ${\Delta}v$
 is probably due to the combined effect of rotational and random 
motion of the absorbing cloud system, plus some wavelength smearing 
because of the extension of the source.
Figure 6 shows the fits to the O {\sc i} 1039 \AA\ absorption line 
and to the N~{\sc i} 1134 \AA\ multiplet. The best fit 
is shown by the thick solid line. It is evident 
that the multi-component fit is much better than the single-component fit
(thin line in Figure 6b). Figure 6a also 
shows an example of a narrow component along one light of sight 
which has a saturated profile (dotted line). 

The heavy element column densities derived with the multi-component 
fit, along with their 
2$\sigma$ error bars are given in Table 2 under the heading log $N_2$(species).
As expected, the derived abundances are larger than in the case of a single 
component fit, by a mean factor of $\sim$ 5. The very large column densities 
derived for C {\sc iii}, S {\sc iii} and Fe {\sc iii} are caused by 
stellar contamination as discussed before. We 
obtain log $N$(O {\sc i})/$N$(H {\sc i})
 = --5.0$\pm$0.3 or [O {\sc i}/H {\sc i}] = --1.5.
It is highly probable that [O/H] $\sim$  [O {\sc i}/H {\sc i}] as 
the ionization potentials of Hydrogen and Oxygen are very close to each other.
 Moreover, the charge exchange cross-section of O {\sc ii} with H {\sc i} 
is large, making it unlikely that Oxygen is more ionized than 
Hydrogen. This appears to be supported 
by the remarkable constancy of [O {\sc i}/H {\sc i}] 
in the local ISM of the Galaxy, independently of 
direction and ionizing flux (Meyer et al. 1998), 
in strong contrast to [N {\sc i}/H {\sc i}] 
which depends on the ionizing conditions. Given that  [O/H] = --5.0,  
the H {\sc i} absorbing cloud 
has a metallicity lower than that of the supergiant H {\sc ii} region 
by a factor of $\sim$ 10. This suggests self-contamination of 
the H {\sc ii} region by heavy elements released during the present burst of 
star formation (Kunth \& Sargent 1986). 
While mixing of these newly formed heavy elements 
appears to have occurred on a scale of $\sim$ 2 kpc as shown by the small 
scatter of the average metallicities of the H {\sc ii} regions in the vicinity 
of the supergiant  H {\sc ii} region (Noeske et al. 2000), 
it has not had time to occur for the whole neutral gas component as  
the H {\sc i}
envelope surrounding the star-forming regions is much more extended 
(its diameter is $\sim$ 19 kpc from the VLA map by Thuan 
et al. 2001). 

\subsection{Modeling}

We attempt next to model the observed heavy element column densities. 
The presence of the higher ionization stages of these elements
implies that the ions reside not exclusively in the H {\sc i} envelope, but 
also in the H {\sc ii} regions around the ionizing stars.
We use the CLOUDY code (Ferland 1996, Ferland et al. 1998; version c90.05)  
to construct a series of photoionized H {\sc ii} region models, and 
select the one which best reproduces 
the optical nebular emission line intensities observed in Mrk 59 (Izotov
et al. 1997). By comparing the predicted column densities 
of the best photoionization model 
with those observed by {\sl FUSE}, we will be able to tell which fraction of 
metals resides in the neutral gas and which fraction is in the ionized gas.   
  
We consider a spherically symmetric ionization-bounded H {\sc ii} region 
model. The calculations 
are stopped in the zone away from the ionizing stars 
where the temperature drops to 2000 K. 
The ionization in this zone is very low and it is taken to be the outer 
edge of the H {\sc ii} region.
Several input parameters need to be set. First, the distance to Mrk 59 is
taken to be 10.7 Mpc. For this distance and using
the aperture-corrected H$\beta$ flux from Guseva et al. (2000),
we derive an H$\beta$ luminosity $L$(H$\beta$) = 1.15 $\times$ 10$^{40}$ 
erg s$^{-1}$, and a number of ionizing photons $N$(Lyc) = 2.45 $\times$ 
10$^{52}$ s$^{-1}$.
We use the Kurucz (1991) stellar atmosphere models and adopt the
effective temperature of the ionizing stellar radiation to be 
 $T_{\rm eff}$ = 50,000 K,
a typical value for low-metallicity high-excitation H {\sc ii} regions. 
For the 
inner radius of the H {\sc ii} region, we adopt $R_{\rm in}$ = 10$^{19}$ cm.
The chemical composition of the H {\sc ii} region is set by the observed 
element
abundances derived from optical spectroscopy of Mrk 59 (Izotov \&
Thuan 1999), except for the carbon abundance. For carbon we adopt log C/O =
--0.4 which is intermediate between the solar value and that for the 
lowest-metallicity BCDs  (Izotov \& Thuan 1999). The adopted abundances are 
shown in Table \ref{Tab3}.

We run the CLOUDY code varying the filling factor $f$ and the electron number
density in order to obtain the best agreement between the predicted and
observed [O {\sc ii}] and [O {\sc iii}] emission lines. 
The best model is found for a filling
factor $f$ = 0.1 and an electron number density $N_{\rm e}$ = 25 cm$^{-3}$.
The optical line intensities of the best model are shown in Table 4, where
we compare them with the observed ones (Izotov et al. 1997).
 There is good general agreement, giving us confidence that the 
photoionization model is correct.
The predicted total hydrogen (neutral and ionized) column density in the 
 H {\sc ii} region is
$N$(H {\sc i} + H {\sc ii}) = 1.88 $\times$ 10$^{21}$ cm$^{-2}$ or 
log $N$(H {\sc i} + H {\sc ii})= 21.27. 
As expected for a H {\sc ii} region model, most of the hydrogen is ionized:
 $N$(H {\sc ii}) = 1.86 $\times$ 10$^{21}$ cm$^{-2}$. The column density
of the neutral hydrogen is two orders of magnitude lower. The model also
predicts a column density of molecular hydrogen made via H$^-$ 
$N$(H$_2$) = 5.78 $\times$ 10$^{11}$ cm$^{-2}$, more than 3 orders of 
magnitude lower than our observational upper limit. 

We now compare the predictions for the column densities of heavy elements with
those derived from the {\sl FUSE} spectrum in the 
more realistic case of multiple velocity component fitting
 (column 5 of Table \ref{Tab2}). The CLOUDY predictions are  
shown in Table \ref{Tab5}. The ones relevant to the {\sl FUSE} observations 
are also given in column 6 of Table \ref{Tab2}. 
In Table \ref{Tab5},  $x$ is the radially averaged ratio of the
ion number to the total number of a particular element, 
e.g., $N$(Fe$^+$)/$N$(Fe).
The predicted log $N$(X) of species X is then derived as 
log $N$(X) = log $N$(H~{\sc i}+H {\sc ii}) + log X$^{+i}$/H -- 
log $x$(X$^{+i}$).
Comparison of columns 5 and 6 in Table \ref{Tab2} reveals that for all 
lines that do not suffer from stellar contamination, the derived 
column densities are larger than the CLOUDY predictions by a mean factor of 
$\sim$ 5 (a factor of 2.5 for O {\sc i}), 
except for the S {\sc iv} ion which has a column density  
smaller than the predicted one, but consistent with it within the errors.
Thus it is clear that there has been previous metal enrichment of the 
neutral ISM of Mrk 59. While the neutral gas is less metal-rich than 
the ionized gas by a factor 10 (section 5.2), the neutral 
gas column densities are larger by a factor of $\sim$ 5 because of the 
larger extent of the neutral as compared to the ionized gas.

\section{The O {\sc vi} and S {\sc vi} stellar lines and their 
P-Cygni profiles }

 Several stellar emission and absorption lines are also seen in the
spectrum of Mrk 59. They are marked in Figure \ref{Fig1} by dotted lines.
The  S {\sc vi} 933.4, 944.5 \AA, O {\sc vi} 1031.9, 1037.6 \AA\ 
lines show redshifted emission associated with blueshifted 
absorption, characteristic of stellar lines with P Cygni profiles.
Such broad P Cygni lines are known to originate in 
stellar winds of O stars, and  
have been detected in several environments 
during previous space missions: in the Galaxy by {\sl Copernicus} 
(Snow \& Jenkins 1977; Morton 1976, 1979; Morton \& Underhill 1977;
Walborn \& Bohlin 1996) and in the Magellanic Clouds during 
the {\sl HUT/Astro2} mission
(Walborn et al. 1995). More recently, studies of O stars with stellar winds
have been undertaken with {\sl FUSE} (Bianchi et al. 2000; 
Fullerton et al. 2000). Although the 
O {\sc vi} and S {\sc vi} stellar lines with
P Cygni profiles have been seen in O stars of all spectral and luminosity
subtypes, they are most prominent and show the strongest redshifted emission
in the hottest and most luminous 
stars: for example, in the O4f star $\zeta$ Pup in the Galaxy  
(Morton 1976; Morton \& Underhill 1977) and
 in some O3 -- O4 stars in the Large and Small Magellanic Clouds
 (Walborn et al. 1995). 
Morton (1976) estimated the mass of $\zeta$ Pup
to be $\sim$ 100 $M_\odot$. 
The presence of the O {\sc vi} and S {\sc vi} P Cygni profiles in the 
{\sl FUSE} spectrum of Mrk 59 thus implies the presence of  
such very massive O stars in the BCD.
 The spectrum does not have, however, the necessary signal-to-noise ratio to 
allow us to carry out a detailed quantitative analysis of the
properties of those stars with stellar winds. 

Additionally, a broad blueshifted C {\sc iii} 1175.7 \AA\ stellar 
absorption line is seen in the spectrum of Mrk 59, with very little 
redshifted emission.  This type of P-Cygni
profile associated with the C {\sc iii} line 
is prominent in the spectra of O and early B supergiants (e.g.,
Walborn \& Bohlin 1996). Unlike the O {\sc vi} and S {\sc vi} 
lines, the C {\sc iii} line is relatively free of contamination 
by other absorption lines, and its width 
allows us to estimate the velocity of the 
stellar wind associated with the massive stars in Mrk 59.
The width as measured from the rest wavelength
to the wavelength of the 
most blueshifted point of the profile is $\sim$ 4 \AA, 
which corresponds to a terminal stellar wind velocity of 
$\sim$ 1000 km s$^{-1}$.
This value is slightly smaller than the terminal velocities of  
$\sim$ 1400 km s$^{-1}$ found for O stars in the Small Magellanic Cloud
 (e.g., Bianchi et al. 2000),
which has a metallicity comparable to Mrk 59.  
 That terminal velocity is on the other hand  
considerably lower than typical wind velocities observed 
for O stars in the Galaxy. For example, Morton (1979)
derived a terminal velocity of 2660 km s$^{-1}$ for $\zeta$ Pup.
However, the terminal stellar wind velocity
 in Mrk 59 is larger than
the velocity of $\sim$ 500 km s$^{-1}$ found by Thuan \& Izotov (1997)
in SBS 0335--052 which has a much lower metallicity (1/41 solar). 
There appears to be a general trend of smaller terminal velocities with 
decreasing metallicities. This trend can be understood since less metals means
less opacity to drive stellar winds and hence lower velocities.

It is very likely that the width
of the blueshifted absorption in the O {\sc vi} P-Cygni profiles 
is similar to
that in the C {\sc iii} line. The expected broad absorption of the two 
O {\sc vi} 1031.9, 1037.6 \AA\ lines 
would then be superimposed on the red wing of the Ly$\beta$ line,
 causing a broad depression in that part of its profile. Figure \ref{Fig1} 
shows indeed that the profile of the Ly$\beta$ line is distinctly asymmetric,
with the red wing being generally at a lower flux level than the blue wing.
We note that this asymmetry of the Ly$\beta$ line is also present 
in the {\sl FUSE} spectrum of other starburst galaxies such as NGC 1705 
(Heckman et al. 2001).

There is no evidence for a velocity shift between the absorption lines arising 
from the gas and the stellar features, so that there is no outflow motion 
from the starburst region.

\section{Summary}

We present new {\sl FUSE} far-UV spectroscopy of the 
metal-deficient ($Z_\odot$/8) cometary Blue Compact
Dwarf (BCD) galaxy Markarian 59 (Mrk 59). We have obtained the following 
results:

1. The very high signal-to-noise of the {\sl FUSE} spectrum allows us 
to study the absorption lines from the nearly entire Lyman series of 
Hydrogen from Ly $\beta$ to Ly $\lambda$ (H {\sc i} 11). The  H {\sc i} lines 
are characterized by narrow cores and broad damping wings, requiring 
a bimodal distribution of H {\sc i} column densities to fit their profiles,
with one set of column densities with values of a few times 10$^{20}$ cm$^{-2}$ 
and the other with values of $\sim$ 10$^{23}$ cm$^{-2}$. This bimodality can be 
understood if the cores of the 
H {\sc i} absorption lines are interstellar in origin while the broad 
damping wings are produced by stellar absorption. 
By fitting the cores of the Lyman series 
lines, we obtain a mean interstellar H {\sc i} column density in 
Mrk 59 of $\sim$ 7 $\times$ 10$^{20}$ cm$^{-2}$.

2. No H$_2$ lines are detected in Mrk 59. We set a 10 $\sigma$ upper limit 
for the total column density of diffuse H$_2$ of 10$^{15}$ cm$^{-2}$. This 
low H$_2$ column density, as compared to the one in the 
 Milky Way which has a similar  
H {\sc i} column density, is due to the combined 
effects of a large UV flux which destroys H$_2$ molecules and of a low 
metallicity resulting in a scarcity 
of dust grains on which to form the molecules. This limit does 
not exclude the possible presence of clumpy H$_2$ in dusty compact 
star-forming regions from which the far-UV photons cannot escape, and which 
{\sl FUSE} cannot see.

3. Many interstellar absorption lines of neutral and singly, doubly and 
triply ionized ions of heavy elements are detected in the {\sl FUSE} 
spectrum of Mrk 59. The most prominent ones are the N {\sc i} 1135.0 \AA,
N {\sc ii} 1084.0 \AA, O {\sc i} 1039.2 \AA,  S {\sc iii} 1012.5 \AA,   
S {\sc iv} 1062.7 \AA\  and  Fe {\sc ii} 1144.9 \AA\ lines. 
We have derived element abundances by modeling the profiles of 
detected absorption lines as the sum of 
multiple profiles with velocity dispersion 
$b$ = 7 km s$^{-1}$ and radial velocity spanning a range of 100 km s$^{-1}$,
some of which can be saturated.  
We find log $N$(O {\sc i})/$N$(H {\sc i})
 = --5.0$\pm$0.3 or [O {\sc i}/H {\sc i}] = --1.5 for the neutral gas, 
about a factor of 10 below the oxygen abundance of the supergiant 
H {\sc ii} region, and implying self-enrichment of the latter.

4. We have used the CLOUDY code to construct a photoionized H {\sc ii} region 
model which best reproduces the optical emission line intensities in 
the supergiant H {\sc ii} region. 
The derived column densities for heavy elements in the neutral gas 
are about 5 times higher than those predicted by CLOUDY for the ionized gas,
implying previous metal enrichment of the H {\sc i} gas. Although the 
neutral gas is a factor of 10 more metal-deficient than the ionized gas,
its heavy element column densities are 5 times higher because of the 
considerably larger extent of the H {\sc i} envelope as compared to the 
supergiant H {\sc ii} region.

5. The  S {\sc vi} 933.4, 944.5 \AA\ and  O {\sc vi} 1031.9, 1037.6 \AA\ 
lines with P-Cygni profiles are seen, originating in stellar winds of 
hot O stars. The terminal stellar wind velocity is $\sim$ 1000 km s$^{-1}$ 
as compared to $\sim$ 2500 km s$^{-1}$ in the Milky Way,
suggesting a trend of smaller terminal velocities with decreasing 
metallicities.

\acknowledgments

The profile fitting was done using the Owens procedure  
 developed by M. Lemoine and the {\sl FUSE} French Team.
T.X.T. thanks the financial support of NASA grant NAG5-8954. He 
acknowledges the hospitality of the Institut d'Astrophysique de Paris and the 
Departement d'Astronomie extragalactique et de Cosmologie at the 
Observatoire de Meudon during his 
sabbatical year. T.X.T. and Y.I.I.
are grateful for the partial financial support of NSF grant AST-96-16863.
Y.I.I. thanks the hospitality of the Astronomy Department of the University of 
Virginia.

\clearpage

%

\begin{deluxetable}{lcc}
\tablenum{1}
\tablecolumns{3}
\tablewidth{0pt}
\tablecaption{Upper limits on the H$_2$ content of Mrk 59
at the $\sim$ 10 $\sigma$ level \label{Tab1}}
\tablehead{
Molecule   & J & $N$(H$_2$)  \\
&&(cm$^{-2}$)}
\startdata
H$_2$ & 0 & $<$ 2 $\times$ 10$^{14}$  \\
      & 1 & $<$ 5 $\times$ 10$^{14}$  \\
      & 2 & $<$ 4 $\times$ 10$^{14}$  \\
      & 3 & $<$ 3 $\times$ 10$^{14}$  \\
      & 4 & $<$ 3 $\times$ 10$^{14}$  \\
      & 5 & $<$ 4 $\times$ 10$^{14}$  \\
\enddata
\end{deluxetable}

\clearpage

%

\begin{deluxetable}{lrrccc}
\tablenum{2}
\tablecolumns{6}
\tablewidth{0pt}
\tablecaption{Heavy element column densities (cm$^{-2}$) in Mrk 59
 \label{Tab2}}
\tablehead{
Species   & $\lambda$(\AA) &$b_1$ (km s$^{-1}$)\tablenotemark{a}
& log $N_1$(species) \tablenotemark{a}
& log $N_2$(species) \tablenotemark{b}& CLOUDY}
\startdata
C {\sc ii}   & 1036.3 &\nodata~  & $>$14.6 &\nodata~             & 15.5 \\
C {\sc iii}  &  977.0 &$\leq$ 35 & 18.4$^{+0.1}_{-0.1}$  & 18.2$^{+0.1}_{-0.1}$
         & 16.8 \\
N {\sc i}    & 1135.0 &40$\pm$20 & 14.0$^{+0.3}_{-0.4}$  & 14.6$^{+0.4}_{-0.3}$
& 13.7 \\
N {\sc ii}   & 1084.0 &30$\pm$10 & 14.2$^{+0.1}_{-0.2}$ & 15.1$^{+2.2}_{-0.8}$
& 14.4 \\
O {\sc i}    & 1039.2 &40$\pm$20 & 15.1$^{+0.3}_{-0.3}$ & 15.8$^{+0.3}_{-0.3}$ 
& 15.4 \\
S {\sc iii}  & 1012.4 &95$\pm$15 & 15.1$^{+0.1}_{-0.1}$ & 19.0$^{+0.1}_{-0.2}$         
& 15.4 \\
S {\sc iv}   & 1062.7 &30$\pm$10 & 14.2$^{+0.1}_{-0.2}$ & 14.9$^{+0.9}_{-0.6}$ 
& 15.3 \\
Fe {\sc ii}  & 1144.9 &40$\pm$15 & 13.9$^{+0.1}_{-0.2}$ & 14.4$^{+0.3}_{-0.3}$ 
& 13.7 \\
Fe {\sc iii} & 1122.5 &105$\pm$10& 15.1$^{+0.1}_{-0.1}$ & 18.9$^{+0.1}_{-0.1}$ 
& 14.1 \\
\enddata
\tablenotetext{a}{Fitting with a single Voigt profile. 
Error bars are 3 $\sigma$ limits.}
\tablenotetext{b}{Fitting with multiple velocity components, 
each with $b_2$ = 7 km s$^{-1}$, and spanning a velocity 
range of 100 km s$^{-1}$.  Error bars are 2 $\sigma$ limits. The 
C {\sc III} 977.0 \AA, S {\sc III} 1012.4 \AA\ and Fe {\sc III} 1122.5 \AA\
lines have high column densities because of stellar contamination.} 
\end{deluxetable}

\clearpage

%
\begin{deluxetable}{lc}
\tablenum{3}
\tablecolumns{2}
\tablewidth{0pt}
\tablecaption{Abundances in Mrk 59 used as input to CLOUDY\tablenotemark{a} \label{Tab3}}
\tablehead{Species   & $N$(species)/$N$(H) } 
\startdata
He\tablenotemark{b} & $-1.08$ \\
C                   & $-4.41$ \\
N                   & $-5.53$ \\
O                   & $-4.01$ \\
Ne                  & $-4.75$ \\
Si                  & $-5.35$ \\
S                   & $-5.55$ \\
Ar                  & $-6.34$ \\
Fe                  & $-5.92$ \\
\enddata
\tablenotetext{a}{From Izotov \& Thuan (1999).}
\tablenotetext{b}{Corresponding to a He mass fraction $Y$ = 0.248
(Izotov \& Thuan 1998).}
\end{deluxetable}

\clearpage

%
\begin{deluxetable}{lcc}
\tablenum{4}
\tablecolumns{3}
\tablewidth{0pt}
\tablecaption{Comparison between the observed and CLOUDY predicted
optical line intensities normalized to H$\beta$ \label{Tab4}}
\tablehead{ Ion  &  Observed\tablenotemark{a} & CLOUDY } 
\startdata
3727 [O {\sc ii}]       &      1.102  & 1.035 \\
3835 H9                 &      0.086  & 0.081 \\
3868 [Ne {\sc iii}]     &      0.449  & 0.502 \\
3889 He {\sc i} + H8    &      0.201  & 0.196 \\
3968 [Ne {\sc iii}] + H7&      0.311  & 0.317 \\
4068 [S {\sc ii}]       &      0.010  & 0.012 \\
4101 H$\delta$          &      0.265  & 0.266 \\
4340 H$\gamma$          &      0.468  & 0.476 \\
4363 [O {\sc iii}]      &      0.089  & 0.082 \\
4471 He {\sc i}         &      0.036  & 0.040 \\
4658 [Fe {\sc iii}]     &      0.005  & 0.004 \\
4740 [Ar {\sc iv}]      &      0.006  & 0.006 \\
4861 H$\beta$           &      1.000  & 1.000 \\
4959 [O {\sc iii}]      &      1.982  & 1.992 \\
5007 [O {\sc iii}]      &      5.802  & 5.750 \\
5271 [Fe {\sc iii}]     &      0.002  & 0.002 \\
5876 He {\sc i}         &      0.106  & 0.109 \\
6300 [O {\sc i}]        &      0.019  & 0.010 \\
6312 [S {\sc iii}]      &      0.017  & 0.019 \\
6363 [O {\sc i}]        &      0.006  & 0.003 \\
6563 H$\alpha$          &      2.819  & 2.833 \\
6584 [N {\sc ii}]       &      0.046  & 0.031 \\
6678 He {\sc i}         &      0.030  & 0.031 \\
6717 [S {\sc ii}]       &      0.095  & 0.074 \\
6731 [S {\sc ii}]       &      0.071  & 0.053 \\
7065 He {\sc i}         &      0.025  & 0.024 \\
7136 [Ar {\sc iii}]     &      0.076  & 0.076 \\
7320 [O {\sc ii}]       &      0.016  & 0.015 \\
7330 [O {\sc ii}]       &      0.014  & 0.013 \\
\enddata
\tablenotetext{a}{From Izotov et al. (1997).}
\end{deluxetable}

\clearpage

%
\begin{deluxetable}{lcc}
\tablenum{5}
\tablecolumns{3}
\tablewidth{250pt}
\tablecaption{CLOUDY predicted column densities \label{Tab5}}
\tablehead{Ion   & log $x$\tablenotemark{a} & log $N$\tablenotemark{b} } 
\startdata
C {\sc ii}~~~~~~~~~~~~~~~~~~~~~~~~~~~~~~   &   $-1.354$ & 15.51 \\
C {\sc iii}  &   $-0.098$ & 16.77 \\
C {\sc iv}   &   $-0.800$ & 16.06 \\
N {\sc i}    &   $-2.023$ & 13.72 \\
N {\sc ii}   &   $-1.367$ & 14.38 \\
N {\sc iii}  &   $-0.104$ & 15.64 \\
O {\sc i}    &   $-1.876$ & 15.39 \\
O {\sc ii}   &   $-1.224$ & 16.04 \\
O {\sc iii}  &   $-0.033$ & 17.23 \\
Si {\sc ii}  &   $-1.256$ & 14.67 \\
Si {\sc iii} &   $-0.324$ & 15.60 \\
Si {\sc iv}  &   $-0.580$ & 15.34 \\
S {\sc ii}   &   $-1.463$ & 14.26 \\
S {\sc iii}  &   $-0.289$ & 15.44 \\
S {\sc iv}   &   $-0.393$ & 15.33 \\
Ar {\sc i}   &   $-2.220$ & 12.71 \\
Fe {\sc ii}  &   $-1.677$ & 13.68 \\
Fe {\sc iii} &   $-1.277$ & 14.08 \\
Fe {\sc iv}  &   $-0.034$ & 15.32 \\
\enddata
\tablenotetext{a}{Radially averaged ratio of the ion number to the total
number of the element.}
\tablenotetext{b}{Column density in cm$^{-2}$.}
\end{deluxetable}

\clearpage


\clearpage


\clearpage


\clearpage


\clearpage


\begin{figure}
\figurenum{5}
\epsscale{1.0}
\plotone{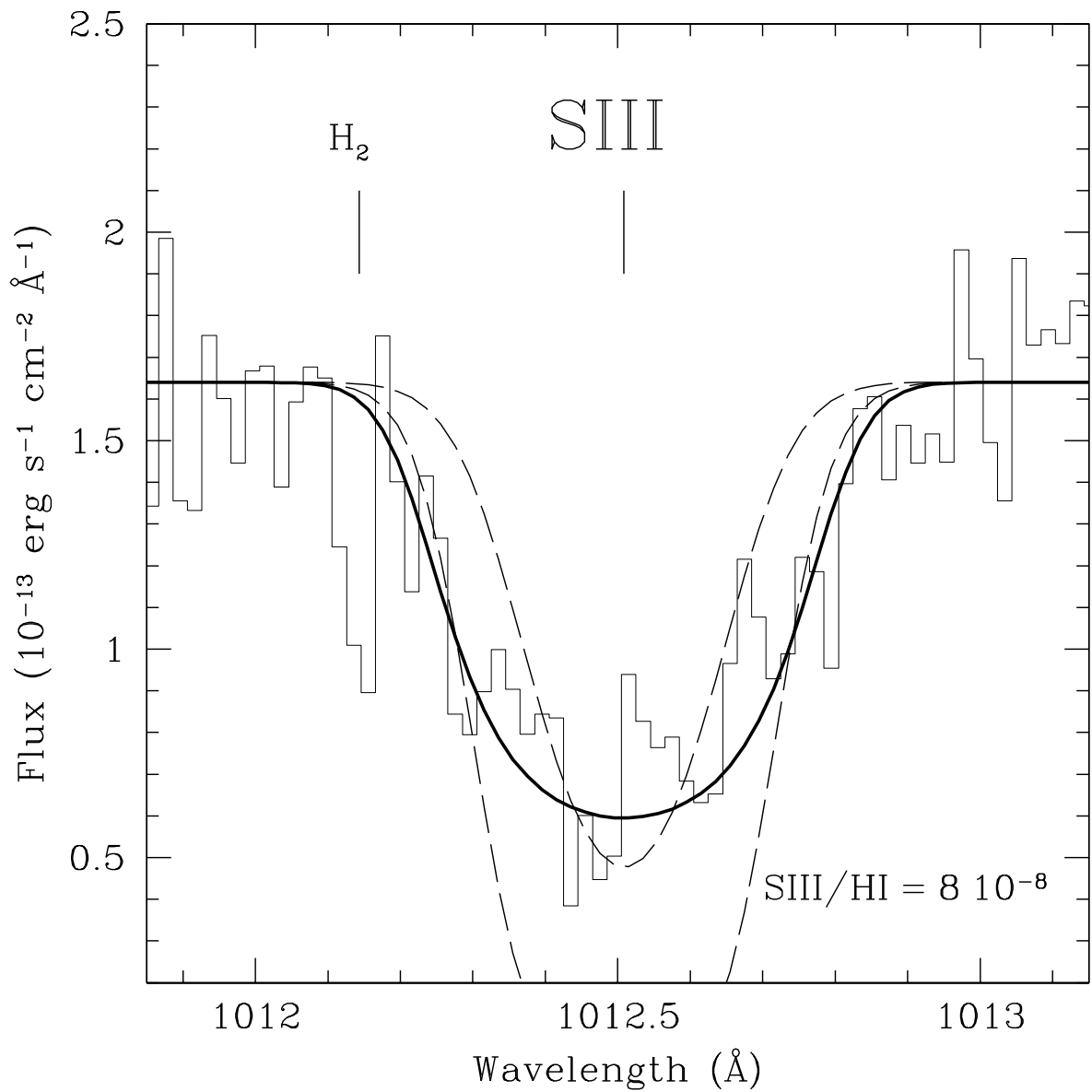}
\caption{\label{Fig5} Profile fitting of the absorption line of ion 
S {\sc iii}. The solid line shows the best fit obtained using the H {\sc i} 
column density distribution shown in Figure \ref{Fig3} and varying the ratio of 
$N$(ion)/$N$(H {\sc i}). The ratio corresponding to the best fit is 
8 $\times$ 10$^{-8}$. The dashed lines show Voigt profile fitting with  
$b$ = 42  km s$^{-1}$ and S {\sc iii} column densities equal respectively to 
8 $\times$ 10$^{14}$ (upper line) and 4 $\times$ 10$^{15}$ cm$^{-2}$ 
(lower line).}
\end{figure}

\clearpage


\begin{figure}
\figurenum{6}
\epsscale{1.0}
\plottwo{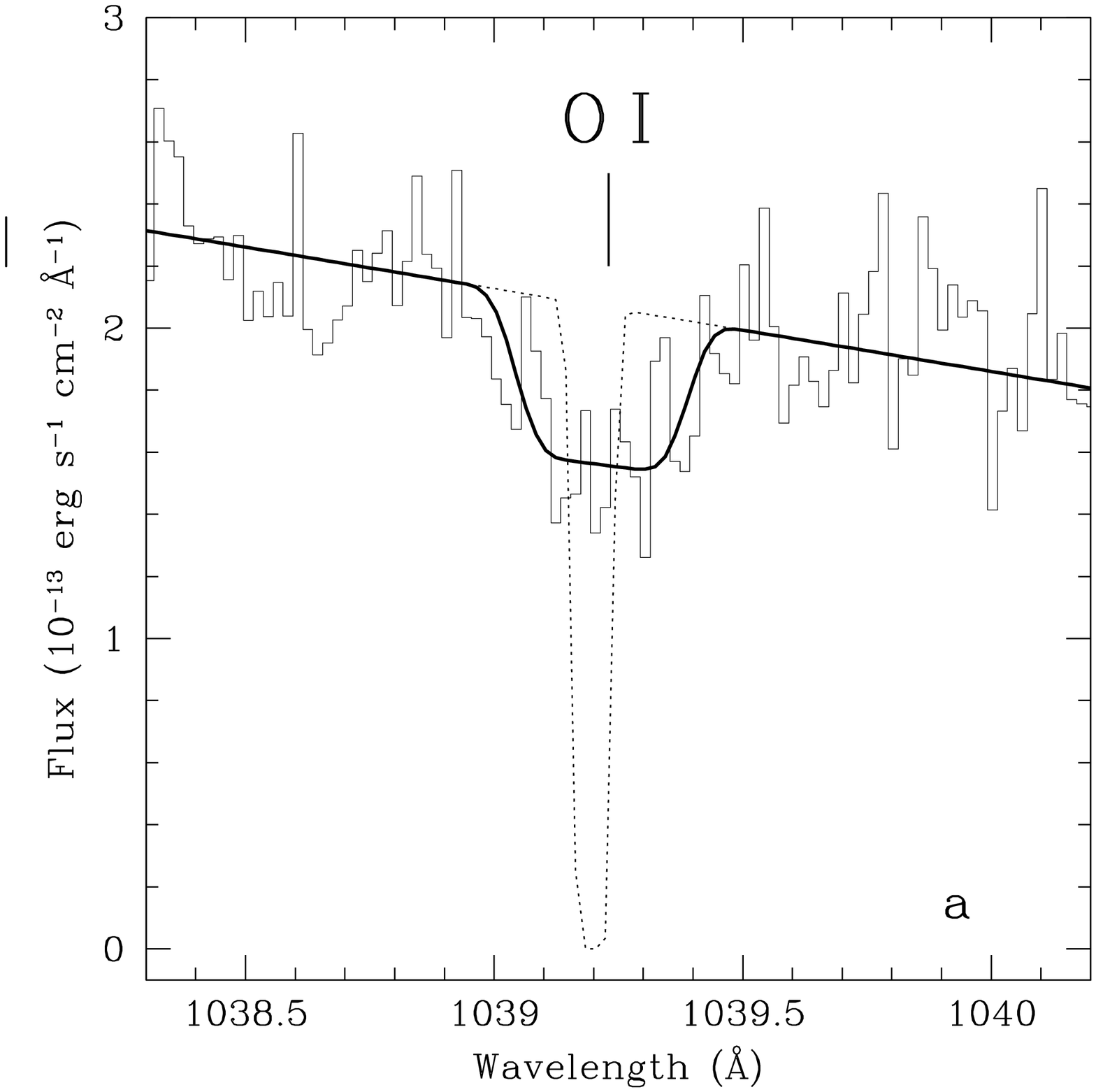}{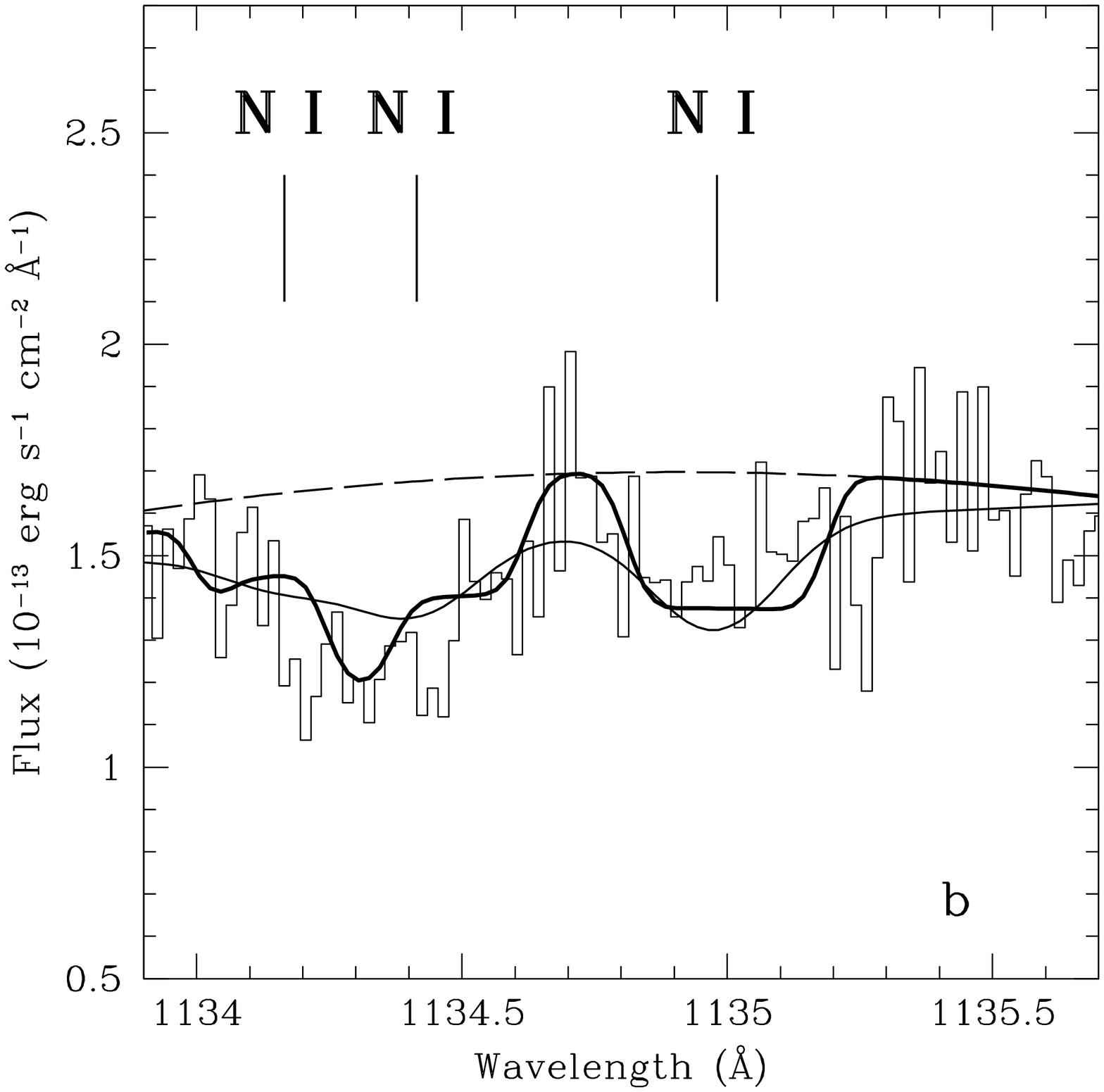}
\caption{\label{Fig6} 
a) Fit to the O~{\sc i} 1039\AA\ absorption line with a
column density $N$(O {\sc i}) = 7 $\times$ 10$^{15}$cm$^{-2}$. 
The fit is obtained by the addition of multiple profiles 
with radial velocities uniformly distributed between 
$v_{\rm Mrk 59}$ -- $\Delta$$v$/2 and $v_{\rm Mrk 59}$ + $\Delta$$v$/2.
$\Delta$$v$ and the $b$ parameter are constrained by the simultaneous fit of 
several lines of O~{\sc i}, N~{\sc i} and Fe~{\sc ii}.
The thick solid line gives the resulting fit to the data with 
$\Delta$$v$ = 100 km~s$^{-1}$ and $b$ = 7 km s$^{-1}$.
The dotted line represents an example of one line of sight with 
$b$ = 7 km s$^{-1}$. With this column density, such an 
individual line of sight is
saturated and goes down to the zero intensity level. 
However the absorption line observed by {\sl FUSE} resulting from the 
addition of many narrower profiles does not go to the 
zero intensity level, although it is broader than the width of the 
instrumental point spread function ($\sim$ 0.1\AA).
b) Fit to the N~{\sc i} 1134\AA\ multiplet with a N~{\sc i} column 
density $N$(N {\sc i}) = 4 $\times$ 10$^{14}$ cm$^{-2}$. 
The thick solid line gives the multi-component fit to the data with 
$\Delta$$v$ = 100 km s$^{-1}$ and $b$ = 7 km s$^{-1}$.
The thin solid line gives the best fit obtained with a simple Voigt profile, 
with $N$(N {\sc i}) = 10$^{14}$ cm$^{-2}$ and $b$ = 40 km s$^{-1}$.
It is clear that the multi-component fit is better
than the single component fit. In the former case, 
the line at 1135\AA\ is saturated and fainter than 
the two other lines at 1134.2\AA\ and 1134.4\AA.}
\end{figure}

\end{document}